# Correcting the charge delocalization error of density functional theory

Emil Proynov and Jing Kong*

Department of Chemistry and Center for Computational Sciences, Middle Tennessee State University, 1301 Main St., Murfreesboro, TN 37130, USA

To the Memory of Dr. Tibor Koritsanszky and the Memory of Dr. Tao Yu

* Correspondence: jing.kong@mtsu.edu

**Abstract**

The charge delocalization error, besides nondynamic correlation, has been a major challenge to density functional theory. Contemporary functionals undershoot the dissociation of symmetric charged dimers $A_2^+$, a simple but stringent test, predict a spurious barrier and improperly delocalize charges for charged molecular clusters. We extend a functional designed for nondynamic correlation to treat the charge delocalization error by modifying the nondynamic correlation for parallel spins. The modified functional eliminates those problems and reduces the multielectron self-interaction error. Furthermore, its results are the closest to those of CCSD(T) in the whole range of the dissociation compared with contemporary functionals. It correctly localizes the net positive charge in $(CH_4)_n^+$ clusters and predicts a nearly constant ionization potential as a result. Testing of the SIE4x4 set shows that the new functional outperforms a wide variety of functionals assessed for this set in the literature. Overall, we show the feasibility of treating charge delocalization together with nondynamic correlation.

Contemporary density functional approximations (DFAs) of density functional theory (DFT), the bread-and-butter tools for computational chemistry, face two major challenges, namely the nondynamic correlation (NDC) and the charge delocalization error. Methods have emerged that make strides in treating the NDC within the single-determinant framework of Kohn-Sham DFT. In particular, Becke has proposed the Becke'05 (B05) functional based on compensating the delocalization of the exact exchange, a basic characteristic of systems with NDC [1,2]. B05 was extended to include the strong correlation occurring in the dissociation limit of covalent bonds, the Becke'13 (B13) method [3,4]. We subsequently modified B13 by applying the adiabatic connection and introducing a model correction for the fractional spin error [5]. This reduced the number of empirical parameters significantly. The method is referred to as Kong-Proynov'16/Becke'13 (KP16/B13) here. B13 and KP16/B13 were shown to recover most of the correlation in cases where the correlation is extremely strong. They reduced the errors for dissociation energies of some covalent-bonds to single digits in kcal/mol [3-5]. KP16/B13 successfully predicted the Mott metal-insulator transition [6], a long standing problem in DFT. Both B13 and KP16/B13 were also shown to be competitive in the prediction of various properties in regular systems [7].

The charge delocalization error has a broad impact on the functional accuracy in various applications: incorrect dissociation limits that involve fractional charge, underestimated reaction barriers and ionization potentials, unrealistic charge distributions, to mention just a few [8-15]. The simplest example of this error is the dissociation of one-electron molecules like $H_2^+$, where most common DFAs yield errors larger than 30 kcal/mol, while HF is exact. The dissociation curves of $H_2^+$ and other $A_2^+$ ('A' typically being a noble gas atom) using various contemporary DFAs show two main common failures: a) Those DFAs undershoot the dissociation, i.e. the calculated energy of the asymptotically fractionally charged species $A^{+0.5}$ is too low, whereas HF overshoots it; b) Most DFAs predict a spurious barrier along the dissociation path whereas HF does not. Another feature of this drawback is that charges are over-delocalized. For instance, when an electron is removed from a cluster of well separated molecules such as methane chains, the positive charge should reside on one of the molecules, but most DFAs would distribute the positive charge across the whole cluster [15]. Similar erratic charge distribution with DFAs is also observed when the geometric symmetry of $A_2^+$ is allowed to break, while HF distributes the charge properly [10].

The charge delocalization error of DFAs has been associated with the self-interaction error (SIE) [10, 16, 17], which occurs when the approximate self-exchange component of a DFA does not cancel exactly the self-Coulomb interaction. HF, on the other hand, is free of one-electron SIE, although it still contains multielectron SIE [10, 17, 18]. Long-range corrected (LRC) DFAs have been a popular remedy for the charge-delocalization error, in which the long-range potential of a DFA is replaced by the HF exchange potential [14, 19, 20]. LRC functionals remove the spurious barrier in the dissociation of $A_2^+$ but still undershoot the dissociation curve and delocalize charges in asymmetric cations [15]. Other DFAs have also been designed in combination with 100% HF exchange in order to alleviate the charge delocalization error [9, 21]. However, increasing the contribution of HF exchange has the danger of playing a 'zero-sum' game. Janesko et al compared many contemporary DFAs for the dissociation limits of $H_2$ and $H_2^+$ among other properties and showed that DFAs that perform well for charge delocalization tend to perform poorly for NDC, and vice versa [9]. Attempts have been made by Yang's group [8, 18] to correct both problems by straightening the exchange-correlation functionals with respect to the effective fractional orbital occupancies. Schemes are also proposed to apply DFT functionals using the HF density post self-consistent-field (SCF) because the HF density is considered more accurate [22, 23].

In this work we assess the KP16/B13 method on the problem of charge delocalization, using $A_2^+$ (A being a noble gas atom), a cluster of well-separated methanes and the SIE4x4 test set [24]. KP16/B13 is designed for the left-right NDC, without explicit consideration of the charge delocalization. Still, it could provide a good foundation for attacking the charge delocalization problem since it includes the full exact exchange. Indeed, the method reduces to HF for one-electron systems and so it is free of one-electron SIE and exact for $H_2^+$. B05 functional [2], a predecessor to B13 and KP16/B13, shows a reduced charge delocalization error for some properties like $H_2^+$ and $He_2^+$ dissociation [25], dipole moments [26] and charge-transfer excitations [12]. In due course of applying KP16/B13 to the aforementioned systems, we find that the parallel-spin NDC component of the model overshoots in some regions, bringing it to about the same or even larger magnitude than the opposite-spin NDC. This is not physically reasonable, because the parallel-spin NDC is a secondary correlation effect by design [2]. We explore here a simple modification that corrects to a large extent this drawback and improves significantly on the charge distribution resulting in reduced multielectron SIE.

KP16/B13 simulates the adiabatic connection (AC) for nondynamic correlation, solving locally the integration over the AC coupling constant $\lambda$ at each reference point with a theoretically motivated $\lambda$-dependent correlation energy density $w_\lambda^c$ [5]:

$$E_{ndc} = \int_0^1 W_\lambda^c \, d\lambda = \int_0^1 \int w_\lambda^c(\mathbf{r}) d^3 r \, d\lambda \tag{1}$$

The resulting correlation energy expression is

$$E_{ndc} = \int \frac{1+e^{bz}}{1-e^{bz}} [1 - \frac{2}{bz} \ln\left(\frac{1+e^{bz}}{2}\right)] u_{ndc} d^3 r \equiv \int q_c^{AC} u_{ndc} d^3 r \tag{2}$$

where $u_{ndc}$ is the nondynamic correlation energy density, and $z$ is a factor of NDC strength [5]:

$$z = \frac{u_{ndc}}{u_{dync}} \tag{3}$$

with $u_{dync}$ being the dynamic correlation energy density at $\lambda = 1$. The factor $z$ is small in regions where the dynamic correlation dominates, and large where the nondynamic correlation dominates. This transition is modulated by the empirical parameter $b$ in Eq.(2). The real space function $q_c^{AC}$ takes into account the correlation kinetic component of NDC.

The form of NDC energy density $u_{ndc}$ used in Eq.(3) is

$$u_{ndc} = (u_{ndc}^{opp} + c\, u_{ndc}^{par}) \tag{4}$$

with the opposite-spin NDC energy density given by [2,3]:

$$u_{ndc}^{opp}(\mathbf{r}) = f_{opp}(\mathbf{r}) \left[ \frac{\rho_\alpha(\mathbf{r})}{\rho_\beta(\mathbf{r})} u_{X\beta}^{ex}(\mathbf{r}) + \frac{\rho_\beta(\mathbf{r})}{\rho_\alpha(\mathbf{r})} u_{X\alpha}^{ex}(\mathbf{r}) \right] \tag{5}$$

where $\rho_\alpha$, $\rho_\beta$ are the electron density per spin direction $\alpha$ and $\beta$ respectively, $u_{X\alpha}^{ex}$, $u_{X\beta}^{ex}$ are the exact-exchange energy density per spin direction $\alpha$ and $\beta$ respectively and $f_{opp}$ is the B05 local nondynamic correlation factor with the modifications described in ref.[27]:

$$f_{opp}(\mathbf{r}) = \min(f_\alpha(\mathbf{r}), f_\beta(\mathbf{r}), 1) \tag{6}$$

$$f_\sigma = \frac{1 - N_{X\sigma}^{eff}(\mathbf{r})}{N_{X(-\sigma)}^{eff}(\mathbf{r})} \tag{7}$$

Here, $N_{X\sigma}^{eff}$ is the B05/B13 relaxed normalization of the exchange hole within a region of roughly atomic size. $f_\sigma$ is a correlation factor measuring how much nondynamic correlation electrons of spin $\sigma$ need to 'borrow' from the opposite spin ($-\sigma$). The min() function in Eq.(6) ensures that the NDC is provided at the level of mutual need between $\alpha$ and $\beta$ electrons. The scaling of the exact exchange by $f_{opp}$ in Eq.(5) adds correlation of opposite spins at $\lambda = 1$ when the exchange hole is delocalized as indicated by $N_{X\sigma}^{eff} < 1$. It accounts for the main NDC in situations such as stretched covalent bonds. $N_{X\sigma}^{eff}$ should not exceed 1 in theory, but in practice it does at some points. It was allowed to be as large as 2 at such points to achieve good accuracy of energy differences [5,27].

The second term in Eq.(4) counts the NDC of electrons with parallel spins. It was found that a secondary compensation for the delocalized exchange is often needed for open-shell systems, with the worst case being the triplet oxygen molecule [1,2]. The factor $c$ in Eq.(4) is an empirical global parameter introduced in KP16/B13 to rescale the original B13 parallel-spin NDC energy density. This energy density has the form [2,5]:

$$u_{ndc}^{par}(\mathbf{r}) = -\frac{1}{2} \sum_\sigma \rho_\sigma(\mathbf{r}) A_{\sigma\sigma}(\mathbf{r}) M_\sigma^{(1)}(\mathbf{r}) \tag{8}$$

where $A_{\sigma\sigma}$ are second-order parallel-spin NDC factors, $M_\sigma^{(1)}$ is the first-order moment of the Becke-Roussel (BR) relaxed exchange hole $\bar{h}_{X\sigma}^{eff}(\mathbf{r},s)$, $s = |\mathbf{r}_2 - \mathbf{r}_1|$ ($n = 1$ in Eq.(9)):

$$M_\sigma^{(n)}(\mathbf{r}) = 4\pi \int_0^\infty s^{n+2} \left| \bar{h}_{X\sigma}^{eff}(\mathbf{r},s) \right| ds \tag{9}$$

$$A_{\sigma\sigma} = \min(A_{1\sigma}, A_{2\sigma}), \quad A_{1\sigma} = \frac{1 - N_{X\sigma}^{eff} - f_{opp} N_{X(-\sigma)}^{eff}}{M_\sigma^{(2)}}, \quad A_{2\sigma} = \frac{D_\sigma}{3\rho_\sigma} \tag{10}$$

$$D_\sigma \equiv \tau_\sigma - \frac{1}{4} \frac{|\nabla \rho_\sigma|^2}{\rho_\sigma}; \quad \tau_\sigma(\mathbf{r}) = \sum_i^{occ} |\nabla \psi_{i\sigma}(\mathbf{r})|^2 \tag{11}$$

Here, $\tau_\sigma$ is the KS kinetic energy density by definition of Becke [28], $\psi_{i\sigma}$'s are the Kohn-Sham occupied orbitals, and $M_\sigma^{(2)}$ is the second-order moment of the Becke-Roussel relaxed exchange hole ($n = 2$ in Eq.(9) ). The second term in the min() function is to avoid the parallel spin correlation in one-electron regions where $A_{2\sigma}$ is identically zero.

The NDC energy density $u_{ndc}$ in Eq.(2) is further modified to decrease the fractional spin error:

$$\bar{u}_{ndc} = u_{ndc}\left(1 + \frac{1}{2}\sqrt{\alpha/\pi}\ \exp(-\alpha/z^2)\sum_\sigma \left(D_\sigma / \rho_\sigma^{5/3}\right)^{1/3}\right) \quad (12)$$

where the parameter $\alpha$ is determined empirically, and $z$ is given by Eq.(3). The well-known function $D_\sigma$ in Eq.(11) [29] is used to eliminate the one-electron SIE. Equation (12) boosts the recovery of nondynamic correlation at the dissociation limit by reducing the fractional-spin error.

The final KP16/B13 energy expression includes the exact exchange $E_x$ (calculated in HF like fashion) and the dynamic correlation energy $E_{dc}$. Here $E_{dc}$ is the $\lambda$-integrated form given by Eqs.(37) and (38) of ref. [3].

$$E_{xc} = E_x + \int q_c^{AC}(\mathbf{r})\bar{u}_{ndc}(\mathbf{r})d\mathbf{r}^3 + E_{dc} \quad (13)$$

The model includes three empirical parameters: $b$ (1.355) in Eq.(2), $c$ (1.128) in Eq.(4) and $\alpha$ (1.128) in Eq.(12), optimized using thermochemical data and fractional spin conditions [5]. Ref.[5] also shows that this functional obeys some know exact conditions stemming from the AC theory [30], as well as the exact finite scaling bound at low electron density limit, and properly scales inhomogeneously with respect to uniform coordinate scaling. On the other hand, its coordinate scaling in high density limit depends on the scaling of $f_{opp}$ in that limit and needs further investigation (See Eq. B15 of Ref.[5]).

While KP16/B13 performs well for various systems with correlation strength from moderate to strong, we find that it does not perform well for the charge delocalization problem. The main issue is that the nondynamic correlation for parallel spins, $u_{ndc}^{par}$ (Eq.(8)), overcompensates the delocalized exchange in this case. When the unpaired electron is delocalized in $A_2^+$, it does not incur significant nondynamic correlation even though the exchange hole is delocalized due to the delocalization of this electron. To limit the contribution of the parallel spin in the multielectron region, we scale here the $A_{2\sigma}$ term in Eq.(10) by a parameter: $A_{2\sigma} \rightarrow kA_{2\sigma}$. We also impose the following local lower bound to the parallel NDC energy density:

$$\frac{1}{2}u_{ndc}^{opp}(\mathbf{r}) \le u_{ndc}^{par}(\mathbf{r}), \text{ for } u_{ndc}^{opp}(\mathbf{r}) < 0 \quad (14)$$

The optimal value of $k$ is found to be 0.07 through a preliminary minimization of the delocalization error of $A_2^+$ molecules (A=Ne, Ar).

The above modification does not impact the performance of the method on singlet systems where strong NDC is often found. It is not expected to impact much open-shell systems like oxygen molecule where the nondynamic correlation of the opposite spins, the main part of the nondynamic correlation, is already substantial. Indeed, the binding energy of oxygen molecule changes only a little with this modification from 4.79 eV to 4.88 eV compared with the experimental value of 5.21eV. The binding energy of the $S_2$ changes from 4.34 eV to 4.29 eV compared to the experimental value of 4.41 eV. Preliminary test on the reaction energies extracted from the MGAE109 dataset [31] using a recently developed script [32] shows that the mean-absolute-deviation (MAD) for these energies changes from 3.91 kcal/mol to 5.69 kcal/mol. Detailed data for this test is provided in the supporting material.

To avoid discontinuities of the SCF potential, we implement the lower bound in Eq.(14) in a smooth fashion, similar to that used in our B05 SCF implementation [27]:

$$u_{ndc}^{par} \leftarrow \left(u_{ndc}^{par} - 0.5 u_{ndc}^{opp}\right)(1+e^{pt})^{-1} + 0.5 u_{ndc}^{opp}, \quad t \equiv \frac{(0.5 u_{ndc}^{opp} - u_{ndc}^{par})}{\sqrt{0.25(u_{ndc}^{opp})^2 - (u_{ndc}^{par})^2}} \quad (15)$$

Here, the smoothing parameter $p$ is adjusted to some large value that still would not affect visibly the SCF convergence pattern ($p = 500$ in this case). Equation (15) provides a smooth cut-off of the parallel-spin NDC energy density $u_{ndc}^{par}$, keeping its values within the physically meaningful range.

MKP16/B13 is compared with KP16/B13, HF, B3LYP and ωB97-X methods for the dissociation of $A_2^+$ molecules (A=He, Ne, Ar), the ionization potential and charge distribution of $(CH_4)_n$ chains (n=1,2,3), and the SIE4x4 set. All the calculations, except with ωB97-X and CCSD(T), are done with ultra-fine unpruned grid with G3LARGE basis set [33] using our home made DFT code xTron which evaluates the exact-exchange energy density efficiently at each grid point [34]. CCSD(T) and ωB97-X calculations are done with the Q-Chem program [35] using G3LARGE basis set. Converged HF, B3LYP, or PBE orbitals are used as the initial guess for KP16/B13 and mKP16/B13 SCF calculations respectively. The dissociation curves are depicted in Fig. 1, the vertical ionization potentials and charge distributions of the methane clusters are listed in Table 1, and the results for the SIE4x4 set are summarized in Fig.2.

The dissociation of $A_2^+$ molecules is an exemplary case of probing the charge delocalization error. Figures 1a-1c show that mKP16/B13 improves significantly over KP16/B13, and performs better than the other methods used in these tests. Its results are also the closest to that of CCSD(T) from near the equilibrium to the dissociation limit. Becke'88 (B88) exchange functional in B3LYP, like other local and semilocal exchange functionals, overestimates the exchange in stretched $A_2^+$ by assuming the delocalized parallel-spin electrons fully local, causing the underestimation of the total energy. The potential of those exchange functionals also drops too fast asymptotically, causing the spurious barrier in the dissociation curves. HF counts the correct number of the parallel-spin electrons locally when $A_2^+$ is well stretched but lacks the correlation energy, and overestimates the total energy. Its potential, however, has the correct asymptotic behavior. LRC functionals such as ωB97-X eliminate the spurious barrier by switching to the HF exchange at long ranges, but still overestimate the exchange effect at short ranges and thus undershoot the dissociation. The result of KP16/B13 is closer to that of CCSD(T) at equilibrium when compared with those of HF, B3LYP and ωB97-X, showing its capability of recovering the majority of the correlation. However, it still undershoots the dissociation with a spurious barrier.

KP16/B13 employs the full HF exchange and it seems surprising that its curves for $Ne_2^+$ and $Ar_2^+$ resemble those of B3LYP (Figs. 1b,1c) albeit with a smaller error. This means the correlation is overestimated by Eq.(13). Between the nondynamic and dynamic correlation, it is the nondynamic one that contributes to the overestimation since the dynamic correlation is of short-range. The nondynamic correlation energy density for opposite spins in this case, $u_{ndc}^{opp}$ in Eq.(5), is very small because all the beta electrons are localized (assuming the delocalized electron is $\alpha$), i.e. $N_{X\beta}^{eff}$ is near one everywhere, and the correlation factors $f_\beta$ and thus $f_{opp}$ are near zero per Eqs.(6) and (7). Thus, the NDC energy density for parallel spins, $u_{ndc}^{par}$ (Eq.(8)), is the main reason for the observed overestimation of the correlation energy. This conclusion is confirmed with numerical calculations. The reason for this overestimation is that $A_{1\alpha}$ in Eq.(10) is larger than zero because $N_{X\alpha}^{eff}$ is smaller than one due to the delocalization of the $\alpha$ electron in the frontier orbital. $A_{2\alpha}$ is near zero for stretched $H_2^+$ and $He_2^+$ because $D_\sigma$ is zero for regions of one-electron per spin, which explains the closeness of HF, KP16/B13 and mKP16/B13 curves for these two systems. $D_\sigma$ is larger than zero for Ne$_2^+$ and Ar$_2^+$ because of the presence of multielectrons per spin in regions where density is significant, which leads to a positive $A_{\alpha\alpha}$ and hence a negative $u_{ndc}^{par}$. Furthermore, $A_{1\alpha}$ increases as the bond is stretched because $N_{X\alpha}^{eff}$ becomes smaller, causing the spurious barrier. The present modification of KP16/B13 with scaling $A_{2\sigma}$ in Eq.(10) and setting a lower bound in Eq.(14) provides a simple but significant correction to $u_{ndc}^{par}$. It adheres to the original motivation of the nondynamic correlation for parallel spin as a secondary correction to that of the opposite spin, and produces better results. Furthermore, its curves for Ne$_2^+$ and Ar$_2^+$ are closer to the zero line than HF in the asymptotic region, correcting some of the multielectron SIE. They are in general close to zero asymptotically, indicating that the method tends to localize charges properly. Overall, the mKP16/B13 curves are the closest to that of CCSD(T), indicating a capability of recovering correlation at all ranges, with reduced multielectron SIE.

The undershooting of the dissociation curve of A$_2^+$ also means an over-delocalization of the charge since A$^{+0.5}$ is predicted to have a lower energy than the average energy of A and A$^+$. This feature is easier to demonstrate through an ensemble of well separated methane molecules with +1 overall charge, without imposing symmetry restrictions in the calculations [15]. The correct picture is that the positive charge is almost exclusively localized on one of the molecules, leading to an almost constant ionization potential of the neutral cluster when the number of molecules of the cluster changes [15]. As shown in Table 1, model functionals that undershoot the dissociation, namely KP16/B13, B3LYP and ωB97-X, predict a distribution of the positive charge between the molecules, and varying ionization potential when the size of the cluster changes. On the other hand, mKP16/B13 and HF correctly localize the positive charge on one of the methanes, and yield an ionization potential essentially independent of the cluster size..

MKP16/B13 is further assessed on the SIE4x4 test set designed for the charge delocalization problem [24]. The set includes four cationic dimers at gradually stretched distances, as shown in Fig.2. It compares mKP16/B13 with a LRC functional (ωB97X-D3) and a double-hybrid functional (DSD-PBEP86). The data for the last two functionals are from ref. [24], where DSD-PBEP86 was shown to have the least MAD among a comprehensive list of contemporary functionals. As shown in Fig.2, the MAD of mKP16/B13 is slightly less than that of DSD-PBEP86. Furthermore, the deviation with mKP16/B13 remains relatively constant throughout the stretching. In contrast, the deviations with the other two methods increase with the stretching, mirroring the A$_2^+$ dissociations discussed earlier. Indeed, mKP16/B13 has the smallest deviation at the largest stretch. This shows that the method is uniquely well balanced in describing molecular charge distribution. Another distinct feature of mKP16/B13 is that its deviations are all negative, whereas the other two methods (and most other methods tested in Ref.[24]) are all positive. Positive deviations lead to charge over-delocalization, as discussed above.

In conclusion, proper treatment of charge distribution, besides the nondynamic correlation, has been a major challenge to DFT. Contemporary functionals are known to undershoot the dissociation of symmetric charged dimer A$_2^+$, a simple but stringent test, and predict a spurious barrier. They over-delocalize charge distributions for charged molecular clusters. The popular long-range correction approach is only a partial remedy. We extend KP16/B13, a functional designed for nondynamic correlation, to treat the charge delocalization. A modification is made to the model to cap appropriately the secondary nondynamic correlation of parallel spins while keeping the nondynamic correlation of opposite spins intact. The modified KP16/B13 eliminates the spurious barrier and the undershooting of the dissociation curve for A$_2^+$, reduces the multielectron self-interaction error, and is the closest to the CCSD(T) estimate in the whole dissociation range, compared to contemporary functionals. It localizes the net positive charge on one molecule for a series of (CH$_4$)$_n^+$ clusters and correctly predicts a nearly constant ionization potential as a result. Testing the modified KP16/B13 on the stringent SIE4x4 set shows that it outperforms contemporary functionals on energetics and charge distribution and contains less multielectron self-interaction error consistently. Overall, we show the feasibility of treating charge delocalization together with nondynamic correlation. We plan to optimize this model further using comprehensive chemical data.

**Acknowledgement**: We enjoyed discussions about DFT with Dr. Koritsanszky and Dr. Tao. This work received support from the National Science Foundation (Grant No. 1665344). We are grateful for Dr. Fenglai Liu's assistance.

| Method | Charges | | | Ionization potential (eV) | | |
|---|---|---|---|---|---|---|
| | M1 | M2 | M3 | $CH_4$ | $(CH_4)_2$ | $(CH_4)_3$ |
| wB97X | +0.492 | +0.003 | +0.505 | 14.287 | 14.259 | 13.933 |
| B3LYP | +0.359 | +0.283 | +0.358 | 14.169 | 13.984 | 12.523 |
| KP16/B13 | +0.505 | +0.476 | +0.024 | 14.224 | 13.590 | 13.619 |
| mKP16/B13 | +0.962 | +0.037 | +0.0008 | 14.310 | 14.267 | 14.247 |
| HF | +0.998 | +0.002 | +0.0005 | 13.321 | 13.447 | 13.293 |

**Table1**. Charge distribution patterns in the $(CH_4)_3^+$ linear cluster with a distance of 5 Å between the monomers. M$i$ is the $i$th molecule on the chain. The ionization potentials are vertical first ionization potentials for the denoted neutral clusters.

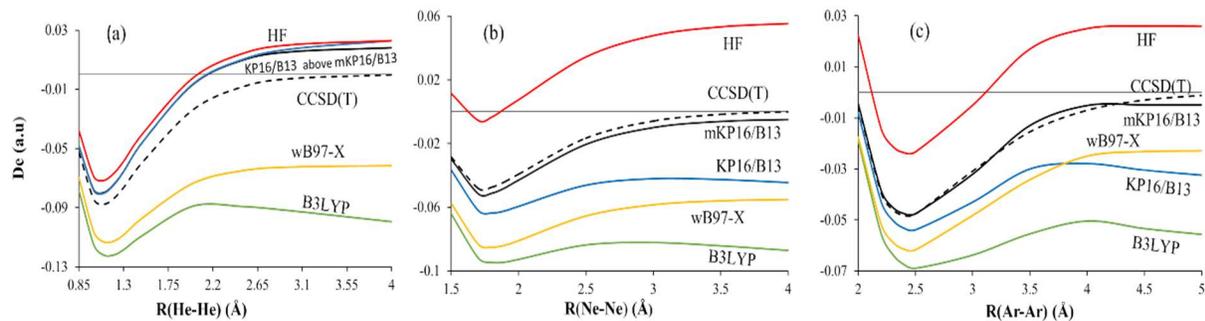

**Figure 1.** Dissociation of $A_2^+$ (A=He,Ne,Ar) with several functionals and CCSD(T).

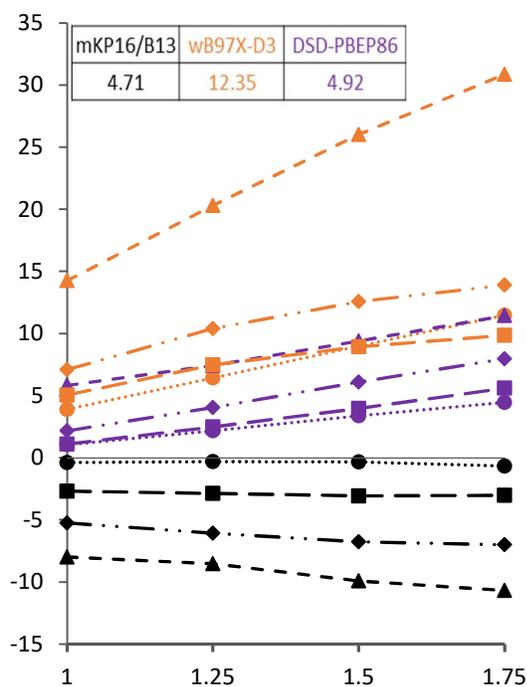

**Figure 2**. Binding energy errors with mKP16/B13, ωB97X-D3 and DSD-PBEP86 for the SIE4x4 data set. Line pattens: circle – $H_2^+$, triangle - $He_2^+$, square – $(NH_3)_2^+$, diamond - $(H_2O)_2^+$. The table on the top shows the MADs for each method and the color used to represent it in the graph. Horizontal axis is the stretching ratio from the equilibrium distance of each cation in the set. Vertical axis is the deviation from the reference value in the binding energy at each distance in kcal/mol. Data for the graph are provided in the Supporting Information.